\documentclass{article}

\usepackage{arxiv}

\usepackage[utf8]{inputenc} 
\usepackage[T1]{fontenc}    
\usepackage{url}            
\usepackage{booktabs}
\usepackage{amsmath}       
\usepackage{amsfonts}       
\usepackage{nicefrac}       
\usepackage{microtype}      
\usepackage{lipsum}         
\usepackage{graphicx}
\usepackage{natbib}
\usepackage{doi}
\usepackage{quiver}
\usepackage{adjustbox}
\usepackage{listings}

\newcommand\numberthis{\addtocounter{equation}{1}\tag{\theequation}}

\tikzcdset{scale cd/.style={every label/.append style={scale=#1},
    cells={nodes={scale=#1}}}}

\title{Music Representing Corpus Virtual: An Open Sourced Library for Explorative Music Generation, Sound Design, and Instrument Creation with Artificial Intelligence and Machine Learning}

\author{\\
	{Christopher Johann Clarke} \\
	In Fulfillment of The National Arts Council Creation Grant\\
	National Arts Council, Singapore\\
	\texttt{chris.johann.clarke@gmail.com} \\
}

\renewcommand{\shorttitle}{MRCV}

\hypersetup{
pdftitle={Music Virtual Corpus: An Open Source Software Suite for Explorative Music Generation, Sound Design, and Instrument Creation},
pdfsubject={},
pdfauthor={Christopher Johann Clarke},
pdfkeywords={Artificial Intelligence, Machine Learning, Education, Music Generation, Sound Design, Instrument Creation, Software, Audio DSP},
}

\begin{document}
\maketitle

\begin{abstract}
	Music Representing Corpus Virtual (MRCV) is an open source software suite designed to explore the capabilities of Artificial Intelligence (AI) and Machine Learning (ML) in Music Generation, Sound Design, and Virtual Instrument Creation (MGSDIC). The software is accessible to users of varying levels of experience, with an emphasis on providing an explorative approach to MGSDIC. The main aim of MRCV is to facilitate creativity, allowing users to customize input datasets for training the neural networks, and offering a range of options for each neural network (thoroughly documented in the Github Wiki). The software suite is designed to be accessible to musicians, audio professionals, sound designers, and composers, regardless of their prior experience in AI or ML. The documentation is prepared in such a way as to abstract technical details, thereby making it easy to understand. The software is open source, meaning users can contribute to its development, and the community can collectively benefit from the insights and experience of other users.
\end{abstract}

\keywords{Artificial Intelligence \and Machine Learning\and Education\and Music Generation\and Sound Design\and Instrument Creation\and Software\and Audio DSP}

\section{Introduction}
\label{sec:intro}
Artificial Intelligence (AI) and Machine Learning (ML) have been areas of active research for several decades, with roots that can be traced back to the 1940s and 1950s \citep{wilson2011affect}. The idea of creating machines capable of emulating human thought and behavior has been a topic of interest since the early days of computing, and the development of AI and ML has been driven by various factors, including improvements in hardware, the availability of larger datasets, and advancements in algorithms and techniques \citep{lu2018brain}.

In recent years, ML has experienced a significant surge of interest due to advancements in deep learning, reinforcement learning, and other techniques. These breakthroughs have led to significant progress in areas such as image and speech recognition, natural language processing, and predictive modeling, among others \citep{howard2019artificial}. The adoption of ML has rapidly spread across numerous industries, including music, where ML is now being used for music generation \citep{kaliakatsos2020artificial}, sound design \citep{miranda1995artificial}, and virtual instrument creation \citep{tahirouglu2020terity}.

Algorithmic music composition has a long history, with early experiments dating back to the 1950s and 1960s. The rise of digital audio workstations (DAWs) in the 1980s and 1990s \citep{jackson2015history} facilitated the creation of more sophisticated music software, which led to the development of algorithmic music composition techniques. These techniques involve the use of computer algorithms to generate musical material, either independently or in collaboration with human composers \citep{alpern1995techniques}.

With the rise of ML techniques in music, the field has experienced an upsurge in interest in using these techniques for music generation, sound design, and virtual instrument creation. Researchers and musicians are exploring the use of ML in music in innovative ways, ranging from the application of supervised and unsupervised learning algorithms to the development of deep neural networks for music analysis \citep{miranda2013readings} and synthesis \citep{roads1985research}. These new techniques have opened up exciting possibilities for musical creativity and exploration, offering new tools for musicians and composers to experiment with and push the boundaries of what is possible in music.

This paper will take a descriptive view on the software library developed, explaining each component part of the library. This serves as a means for further development. As a companion aid to the documentation on the Github, this will highlight the formal structure of the library, and the reasoning behind the design choices made. The paper will also discuss the future development of the library, and the potential for further research in the field of AI and ML in music. The potential implications of this software are outside of the scope of the discussion herein.

\section{Descriptive View of the Software Library}
\label{sec:descriptive}
\subsection{Preface}
\label{subsec:preface}
In order to provide formal description of the various neural networks, this subsection is dedicated to expressing some of the language and notation used in the sections ahead. Firstly, a dense neural network can be described thusly:
\begin{equation}
	\label{eq:dense}
	\hat{y} = \mathbb{M}(x) = \sigma(Wx+b)\circ \dots \circ \sigma(Wx_0 + b)
\end{equation}
Where $\mathbb{M}$ represents the model, which is composed ($\circ$ denotes function composition) of multiple layers. Each layer $\sigma(Wx+b)$ has an activation function $\sigma$, a weight value $W$, an input $x$, and a bias value $b$. $x_0$ is defined as the first input to the model, and $\hat{y}$ is the output of the model.

In most cases, the input $x$ comes from feature extraction. In the case of audio, an example of this is an FFT of the signal at a given time frame. This can be represented as such, borrowing from equation 1:
\begin{equation}
	\label{eq:feature}
	\hat{y} = \mathbb{M}(\mathcal{F}(x)) = \sigma(W(x)+b)\circ \dots \circ \sigma(W\mathcal{F}(x_0) + b)
\end{equation}
Where $\mathcal{F}$ is the feature extraction method used. From Equation \ref{eq:feature}, this feature extraction method is only applicable to the first input of the network.

In order to efficiently notate a model's parameters, such as layer width (number of neurons per layer) and layer count (number of layers), the following shorthand notation is used:
\begin{equation}
	\mathbb{M} = \Bigg[\sum^{(L,\mathbf{p})}_0\Bigg]
\end{equation}
Where $L$ is the layer count, and $\mathbf{p}$ is the layer width. For example, a model with 3 layers, with 2 neurons per layer, can be notated as such:
\begin{equation}
	\mathbb{M}_{\textit{layers}=3,\textit{width}=2}= \Bigg[\sum^{(3,2)}_0\Bigg]
\end{equation}
This notation is used throughout the paper to describe the various models and submodels used.
\subsection{Motivations in Explorative Searching}
Generally, this software library makes use of the trained latent representations of the dataset as a sort of pseudo-random generator. However, this pseudo-randomness maintains some sort of congruence to the training dataset. A latent representation in a neural network refers to the hidden layers that capture a compressed, abstract representation of the input data. The goal of this layer is to extract meaningful and relevant features from the input to the layer and transform these input data into a differently compact or expressive compressed form.

During the training process, the neural network learns to map the input data onto this latent representation by adjusting the weights of the connections between the layers. This process is known as feature extraction and is critical to the success of many machine learning tasks, such as image classification, natural language processing, and music generation.

The term "latent" refers to the fact that the representation is not directly observable in the input data but is inferred from the patterns and relationships within the data. This compressed representation can be thought of as a high-dimensional abstraction of the input data that captures the essential characteristics of the data, while discarding irrelevant details.

Once the neural network has learned a good latent representation of the input data, it can be used for a variety of tasks, such as generating new data that is similar to the input data or classifying the input data into different categories. Overall, latent representations are a powerful tool for neural networks and are a key factor in their ability to learn complex relationships within data.

However, when used in this case, the latent representations serve as a means of mixing multiple characteristics into one. For example, the latent representation of a saxophone note can be mixed with the latent representation of a flute note, resulting in a hybrid sound. However, this resulting hybrid sound might bear almost no similarity to the input sound. In essence, this is a form of ``using neural networks the wrong way'' or neural network bending (derivate of circuit bending). This is the basis of the software library, and the following sections will explain the various components of the library.

The table below shows the various sections of the library, this section will explain each section in detail. The sections are as follows:
\begin{table*}[h]
	\centering
	\caption{Music Representing Corpus Virtual (MRCV)}
	\label{tab:sections}
	\begin{tabular}{ll}
		\toprule
		Section          & Description                                                             \\
		\midrule
		Neural Network 1 & Music Generation                                                        \\
		Neural Network 2 & Sampler Instrument Procedural Generation (Sound Design)                 \\
		Neural Network 3 & Realtime Audio-to-Audio Inferencing (VST/AU plugin)                     \\
		Neural Network 4 & Neural Wavetable Generation through Mel-frequency Cepstrum Coefficients \\
		Genere           & Procedural Score Generation                                             \\
		Audio Dataset 1  & Saxophone Ordinario Dataset                                             \\
		Audio Dataset 2  & Saxophone Multiphonic Dataset                                           \\
		Audio Dataset 3  & Piano Dataset                                                           \\
		MIDI Dataset     & MAESTRO Dataset                                                         \\
		\bottomrule
	\end{tabular}
\end{table*}

\subsection{Neural Network 1: Music Generation}
\label{sec:nn1}
\subsubsection{Introduction}
The first neural network offered is a Multiple-In-Multiple-Out (MIMO) Neural Network model \citep{ayomoh2012neural} or MixMo Neural Network model \citep{rame2021mixmo}. Since the input and output of this neural network is MIDI, the neural network will have to be able to ingest the correct format for each column of data in the MIDI dataset. The chosen format of data is a list containing:
\begin{description}
	\item[Onset Time] Initial time value when note is played
	\item[Duration] Time value when note is released (offset time - onset time)
	\item[Pitch] MIDI pitch value
	\item[Velocity] MIDI velocity value corresponding to loudness (and sometimes timbre)
\end{description}
This neural network only makes use of this four input variables. The given task, and by extension the loss function of the network, is to predict the next note in the series of notes. Functionally, we can express that as such:
\begin{align*}
	\label{eq:nn1}
	\mathcal{L} = f(\textit{Note}_{t+1} - \textit{Note}_{t})
	\quad \mathcal{L}_\textit{components}
	\begin{cases}
		\textit{Onset Time} \\
		\textit{Duration}   \\
		\textit{Pitch}      \\
		\textit{Velocity}
	\end{cases}
	\numberthis
\end{align*}
Since there are 4 separate components within the loss function, it will be hard to linearly combine these various components of loss to derive a single loss. Instead, the neural network can be made to predict 4 separate values. This is done by having 4 separate output layers, each with their own network and loss function. The loss function for each output layer is the mean squared error (MSE) loss function. The MSE loss function is defined as:
\begin{equation}
	\label{eq:mse}
	\mathcal{L}_\textit{MSE} = \frac{1}{n}\sum_{i=1}^{n}(y_i - \hat{y}_i)^2
\end{equation}
\subsubsection{Model}
Conceivably, this leaves us with the motivation to construct a network that makes use of multiple inputs and produces multiple outputs. This is further informed by the fact that compositionally speaking. The various components of a note are not independent of each other. There might be interactions between each component of the note, for example, the onset time of a note might be dependent on the pitch of the previous note. With this understanding, a neural network can be constructed as such:
\begin{align*}
	\textit{sub}\mathbb{M}_{\textit{pitch,duration}} = \Bigg[\sum^{(L,\mathbf{p})}_0\Bigg](x_{\textit{pitch}}\cup x_{\textit{duration}})       \quad\quad\quad\quad
	\textit{sub}\mathbb{M}_{\textit{pitch,onset}} = \Bigg[\sum^{(L,\mathbf{p})}_0\Bigg](x_{\textit{pitch}}\cup x_{\textit{onset}})              \\
	\textit{sub}\mathbb{M}_{\textit{pitch,velocity}} = \Bigg[\sum^{(L,\mathbf{p})}_0\Bigg](x_{\textit{pitch}}\cup x_{\textit{velocity}})     \quad
	\textit{sub}\mathbb{M}_{\textit{duration,velocity}} = \Bigg[\sum^{(L,\mathbf{p})}_0\Bigg](x_{\textit{duration}} \cup x_{\textit{velocity}}) \\
	\textit{sub}\mathbb{M}_{\textit{onset,velocity}} = \Bigg[\sum^{(L,\mathbf{p})}_0\Bigg](x_{\textit{onset}}\cup x_{\textit{velocity}})     \quad\quad
	\textit{sub}\mathbb{M}_{\textit{duration,onset}} = \Bigg[\sum^{(L,\mathbf{p})}_0\Bigg](x_{\textit{duration}}\cup x_{\textit{onset}})        \\
	\forall \hat{y} = \{\hat{y}_{\textit{pitch}}, \hat{y}_{\textit{onset}}, \hat{y}_{\textit{duration}},\hat{y}_{\textit{velocity}}\} = \Bigg[\sum^{(L,\mathbf{p})}_0\Bigg]_i\bigcup_{i=1}^{n_{\textit{sub}\mathbb{M}}}\genfrac(){0pt}{0}{n_{\textit{sub}\mathbb{M}}}{3_{\forall x}}\textit{sub}\mathbb{M}_i\quad\quad\quad\quad\quad
	\numberthis
	\label{eq:nn1_expand}
\end{align*}

\vspace{8mm}
For further clarification, the notation used in Equation \ref{eq:nn1_expand} is described in the figure below:
\begin{figure*}[h]
	\centering
	\[\begin{tikzcd} [scale cd=0.8]
			&& Pitch & Onset & Duration & {\textit{Velocity}} \\
			& \bigcup\bullet & \bigcup\bullet & \bigcup\bullet & \bigcup\bullet & \bigcup\bullet & \bigcup\bullet \\
			{\textit{sub}\mathbb{M}_i} & {\Bigg[\sum^{(L,\mathbf{p})}_0\Bigg]} & {\Bigg[\sum^{(L,\mathbf{p})}_0\Bigg]} & {\Bigg[\sum^{(L,\mathbf{p})}_0\Bigg]} & {\Bigg[\sum^{(L,\mathbf{p})}_0\Bigg]} & {\Bigg[\sum^{(L,\mathbf{p})}_0\Bigg]} & {\Bigg[\sum^{(L,\mathbf{p})}_0\Bigg]} \\
			\\
			{\genfrac(){0pt}{0}{n_{\textit{sub}\mathbb{M}}}{3_{\forall x}}} \\
			&& {\Bigg[\sum^{(L,\mathbf{p})}_0\Bigg]} & {\Bigg[\sum^{(L,\mathbf{p})}_0\Bigg]} & {\Bigg[\sum^{(L,\mathbf{p})}_0\Bigg]} & {\Bigg[\sum^{(L,\mathbf{p})}_0\Bigg]} \\
			&& {\hat{y_{\textit{pitch}}}} & {\hat{y_{\textit{onset}}}} & {\hat{y_{\textit{duration}}}} & {\hat{y_{\textit{velocity}}}}
			\arrow[from=1-3, to=2-2]
			\arrow[from=1-3, to=2-3]
			\arrow[from=1-3, to=2-4]
			\arrow[from=1-4, to=2-2]
			\arrow[from=1-5, to=2-3]
			\arrow[from=1-6, to=2-4]
			\arrow[from=1-4, to=2-5]
			\arrow[from=1-5, to=2-5]
			\arrow[from=1-4, to=2-6]
			\arrow[from=1-6, to=2-6]
			\arrow[from=1-5, to=2-7]
			\arrow[from=1-6, to=2-7]
			\arrow[from=2-3, to=3-3]
			\arrow[from=2-4, to=3-4]
			\arrow[from=2-5, to=3-5]
			\arrow[from=2-6, to=3-6]
			\arrow[from=2-7, to=3-7]
			\arrow[from=2-2, to=3-2]
			\arrow[from=3-2, to=6-3]
			\arrow[from=3-3, to=6-3]
			\arrow[from=3-4, to=6-3]
			\arrow[from=3-5, to=6-4]
			\arrow[from=3-2, to=6-4]
			\arrow[from=3-6, to=6-4]
			\arrow[from=3-3, to=6-5]
			\arrow[from=3-5, to=6-5]
			\arrow[from=3-7, to=6-5]
			\arrow[from=3-4, to=6-6]
			\arrow[from=3-6, to=6-6]
			\arrow[from=3-7, to=6-6]
			\arrow[from=6-3, to=7-3]
			\arrow[from=6-4, to=7-4]
			\arrow[from=6-5, to=7-5]
			\arrow[from=6-6, to=7-6]
		\end{tikzcd}\]
	\caption{The structure of the Neural Network in Equation \ref{eq:nn1_expand}. With each input parameter connected to different submodels within the neural network. These submodels are then combined differently to produce the output parameters.}
	\label{fig:nn1_expand}
\end{figure*}
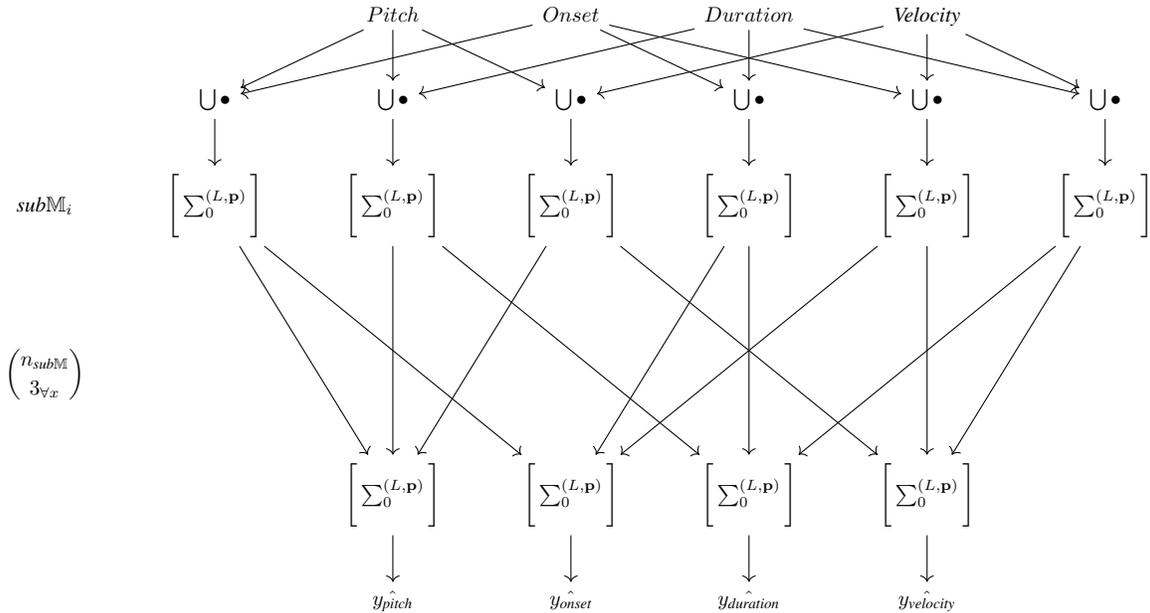

Clarifying the notation used in Equation \ref{eq:nn1_expand}: The $\cup$ operator describes any differentiable operator, which is inclusive of but not limited to the cardinal operators of $+,-,\times,\div$. Specifically, the operator used in the deployed version of the system is the \textit{concat} operator:
\begin{align*}
	a = [1,2,3]                          \\
	b = [4,5,6]                          \\
	\textit{concat}(a,b) = [1,2,3,4,5,6] \\
	\numberthis
\end{align*}
\subsubsection{Dataset}
This model is deployed with the MAESTRO (MIDI and Audio Edited for Synchronous TRacks and Organization) dataset \citep{hawthorne2018enabling}, which is a collection of about 200 hours of virtuosic piano performances captured with fine alignment (~3 ms) between note labels and audio waveforms. The recorded performances are MIDI recordings of virtuoso pianists performing of Yamaha Disklaviers. The repertoire is mostly classical (common practice period), including mostly composers from the 17th to early 20th century.

In the code, a lookup function has been made to construct the dataset. The lookup function takes in an array of strings and parses these strings to return a MIDI dataset with the corresponding composers. An example of this is as such.

\begin{lstlisting}
	midi_data = get_data_for_composer(main_data, ["Handel", "Medtner"])
	# returns a dataset that has only the MIDI files of Handel and Medtner
	# in each column the dataset contains:
	startTimes = midi_data[:, 0]
	endTimes = midi_data[:, 1]
	pitches = midi_data[:, 2]
	velocities = midi_data[:, 3]
	durations = midi_data[:, 4]
\end{lstlisting}

\subsubsection{Output of the Model}
Here is one example of the output of the model, represented as images of piano roll. The x-axis represents time and the y-axis represent pitch height (in midi note numbers). This model was trained on a dataset that comprised of MIDI recordings of Handel and Medtner.

\begin{figure}[h]
	\centering
	\includegraphics[width=\columnwidth]{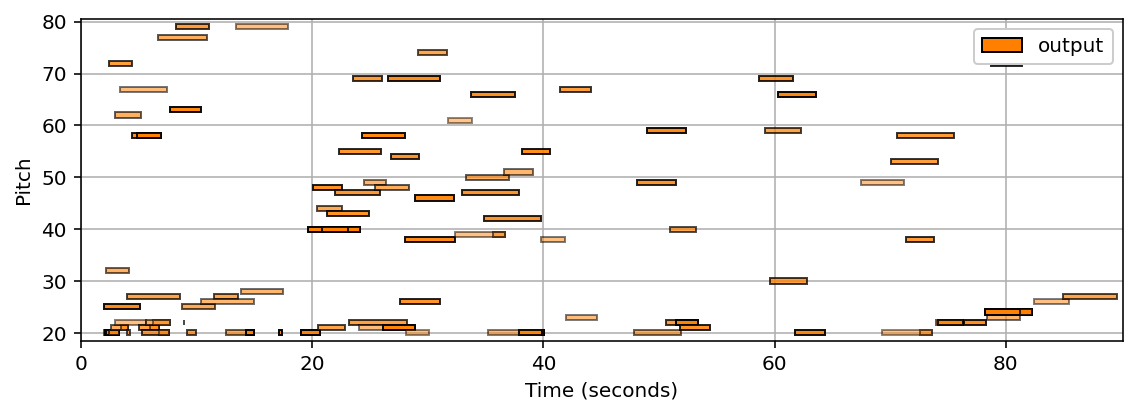}
	\caption{Output of the model. The x-axis represents time and the y-axis represents pitch height (in midi note numbers).}
	\label{fig:outputplot_nn1}
\end{figure}

\subsection{Neural Network 2: Sound Design}
\subsubsection{Introduction}
Continuing with the theme of misusing neural networks, Neural Network 2 is designed to be a sound design tool. The model takes in a sample of an audio at time $t-1$ and is asked to predict the sample at time $t$. An aspect that will be elaborated on further is the potential of mixing datasets. The model is deployed with the Saxophone Multiphonics Dataset as its default training data.
\subsubsection{Model}
The model is a simple dense-layered neural network with a dropout layer after every layer. The dropout layer will be notated as $d$ in the model. Formally, we can refer to the mixture of Datasets $\mathcal{D}$ as:
\begin{equation}
	\mathcal{D} = \bigcup \  ( \{ D_1, D_2, \dots, D_n \} )
\end{equation}
since the model is exposed to a mixture of datasets, the latent representations learnt by the model adhere to these datasets. The model will try its best to generate outputs that adhere to the latent patterns of these datasets.
\begin{align*}
	\hat{y} = \Bigg[\sum^{(L,\mathbf{p})}_0(d)\Bigg](x_{\mathbb{B} \dots \mathbb{B}*})
	\numberthis
\end{align*}
The model has certain network architecture parameters that are exposed to the user. In this case, layer count $L$ and layer width $p$ are exposed to let the user design the model architecture. $\mathbb{B} \dots \mathbb{B}*$ refers to the block size of the input and output. More specifically, if the block size were 4, then the input would be $x_{t-3},x_{t-2}, x_{t-1}, x$, and the next input would be $x_{t+1}, x_{t+2}, x_{t+3}, x_{t+4}$. The output of the model should be trying to return predicted $x_{t+1}, x_{t+2}, x_{t+3}, x_{t+4}$, and $x_{t+5}, x_{t+6}, x_{t+7}, x_{t+8}$ respectively. For further clarification, this is the model architecture in the case of $L=2$ and $p=4$, with 1 input and 1 output from the dataset:
\begin{align*}
	\hat{y} = \Bigg[\sum^{(2,4)}_0(d)\Bigg](x_{t-3},x_{t-2}, x_{t-1}, x) \\
	\hat{y} \approx x_{t+1}, x_{t+2}, x_{t+3}, x_{t+4}
	\numberthis
\end{align*}
Graphically, this can be represented as:

\vspace{8mm}
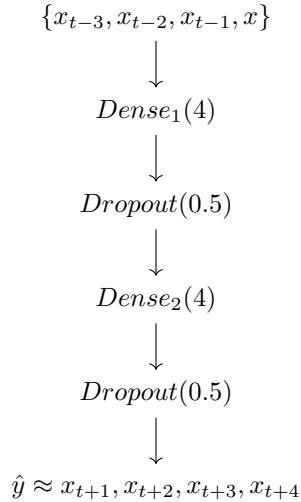
\begin{figure*}[h]
	\centering
	\[\begin{tikzcd} [scale cd=1]
			{  \{ x_{t-3},x_{t-2}, x_{t-1}, x \}} \\
			{Dense_1(4)} \\
			{Dropout(0.5)} \\
			{Dense_2(4)} \\
			{Dropout(0.5)} \\
			{\hat{y} \approx x_{t+1}, x_{t+2}, x_{t+3}, x_{t+4}}
			\arrow[from=1-1, to=2-1]
			\arrow[from=2-1, to=3-1]
			\arrow[from=3-1, to=4-1]
			\arrow[from=4-1, to=5-1]
			\arrow[from=5-1, to=6-1]
		\end{tikzcd}\]
	\caption{The structure of the Neural Network 2.}
	\label{fig:nn2_expand}
\end{figure*}

The model will then be asked to predict $M$ number of blocks. Realistically, each block can be imagined as an audio file containing 44100 samples. Thus the final outputs of the model will resemble a batch of files of length (block size) with $M$ number of files.
\begin{equation}
	\{ \hat{Y} \} = \{ \hat{y}_1, \hat{y}_2, \dots, \hat{y}_M \}
\end{equation}
These files are then passed to a script that will procedurally generate a Decent Sampler Instrument \citep{Shopdece2:online}, which is a sampler instrument that can be used in a Digital Audio Workstation (DAW). The script will procedurally generate a sampler instrument that will play the files in the order of the output. The sampler instrument will be generated with the following user accessible parameters: Attack, Decay, Sustain, and Release.
\subsubsection{Dataset}
The Saxophone Multiphonic Dataset (SMD) is a dataset of saxophone multiphonics. This dataset features a recording of all the multiphonics listed in the saxophone multiphonics book by \citet{kientzy1982sons}. SMD features PCM recordings at a sampling rate of 44.1 kHz. In total, there are 228 audio files.
\subsubsection{Exploratory Sound Design: Output of the Model}
\begin{figure}[h]
	\centering
	\includegraphics[width=0.6\columnwidth]{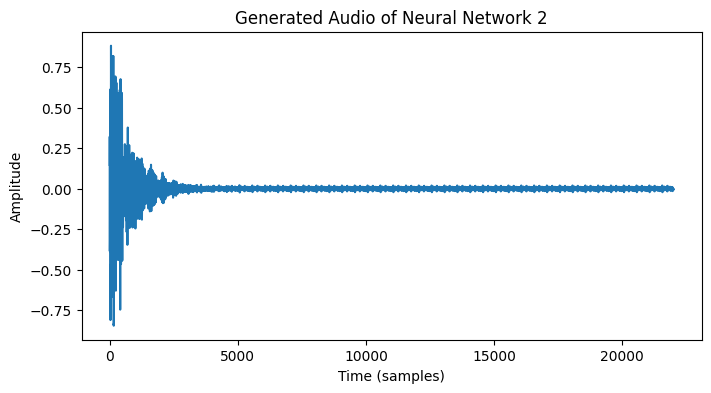}
	\caption{Output of the model. The x-axis represents time and the y-axis represents pitch height (in midi note numbers).}
	\label{fig:outputplot_nn2}
\end{figure}

To highlight the aspect of exploratory sound design, one of the outputs of the model generated when trained on the SMD dataset resembled an electronic hi-hat from a drum machine. This provides insight into the potential of using neural networks as a sound design tool. The output of the model is shown in \autoref{fig:outputplot_nn2}.

\subsubsection{Model Exposed Parameters: Code Example}
This section demonstrates the ease of use of the code when generating a model for this process. The user is able to select layer count $L$, layer width $p$, and block size $\mathbb{B} \dots \mathbb{B}*$.
\begin{lstlisting}
	BLOCK_SIZE = 10
	model = Creator().createDenseModelForNeuralNet2(
                                                        2, 
                                                        32, 
                                                        BLOCK_SIZE, 
                                                        "tanh")
\end{lstlisting}

This code will generate a model with 2 layers, 32 neurons per layer, and a block size of 10.
\subsection{Neural Network 3: Audio Plugin Creation}
\subsubsection{Overview}
There has been numerous neural audio plugins developed by research teams working on various tasks, such as Virtual Analog Modelling, and Neural Methods for Digital Signal Processing, and Neural Audio Synthesis. Neural audio plugins are audio effects plugins that are based on machine learning models, particularly neural networks, for processing audio signals. These plugins can be used to create new sound effects or to modify existing audio in various ways. One common application of neural audio plugins is in Virtual Analog Modelling (VAM) of audio hardware. VAM involves creating a digital model of an analog audio device, such as a vintage synthesizer, to reproduce its sound characteristics digitally. Neural audio plugins have shown promise in improving the accuracy and realism of VAM by modeling the complex nonlinearities and dynamics of analog circuits more accurately than traditional DSP methods.

Another area where neural audio plugins have been applied is in the development of new audio effects, such as distortion, filtering, and time stretching. Neural networks can learn complex mappings between input and output audio signals, which can be used to create novel audio effects that are difficult or impossible to achieve using traditional DSP techniques.

In recent years however, generating high-quality audio waveforms still remains a difficult task, and existing models either require significant computational power, low sampling rates, are difficult to control, or have limited signal generation capabilities. Examples such as Realtime Audio Variational autoEncoder (RAVE) \citep{caillon2021rave} offer both fast and high-quality audio waveform synthesis. The model uses a two-stage training procedure, with representation learning and adversarial fine-tuning, and allows direct control between reconstruction fidelity and representation compactness. Additionally, the model can generate 48kHz audio signals while running 20 times faster than real-time on a standard laptop CPU. The authors evaluate the model's synthesis quality using both quantitative and qualitative subjective experiments and demonstrate its superiority compared to existing models.

This sort of complexity, however useful, is outside of the remit of what MRCV is offering. Instead, MRCV aims to put forth an extremely simple pipeline for new users to train their models and generate new sounds in realtime. The model offered here can be thought of as an audio-to-audio model, wherein, a sample (up to 8 samples of memory) is offered to the model to predict the next time step. This is a very simple model that can be trained on a laptop and used in realtime. The model is also very easy to use, with only a few parameters that can be tweaked to generate new sounds. The model is also very fast, and can generate high sample rate audio in realtime on a standard laptop CPU.

\subsubsection{Model}
The model offered here is a Gated Recurrent Unit (GRU) based model where the user can decide on the number of GRUs and their activation function. There is also a scaling parameter that allows the user to increase the parameters of the model.

GRUs are a type of Recurrent Neural Network (RNN) that is widely used in sequential data processing tasks, such as language modeling, speech recognition, and time series prediction. Unlike traditional RNNs, which suffer from the vanishing gradient problem \citep{hochreiter1998vanishing} that hinders their ability to capture long-term dependencies in data, GRUs are designed to mitigate this issue.

At the core of a GRU is a gating mechanism that selectively updates and outputs information at each time step. The gating mechanism consists of two gates: an update gate and a reset gate. The update gate controls how much of the previous hidden state should be preserved and how much of the new input should be incorporated into the new hidden state. The reset gate determines how much of the previous hidden state should be forgotten and how much of the new input should be used to reset the hidden state.\citep{chung2014empirical} The update and reset gates are implemented using sigmoid and element-wise multiplication operations, which are differentiable and allow for end-to-end training of the GRU using backpropagation. The output of a GRU at each time step is a combination of the current hidden state and the input, which are both passed through a non-linear activation function.

Compared to other types of RNNs, GRUs have a simpler architecture that requires fewer parameters to train, making them faster and more memory-efficient. This makes them well-suited for real-time applications where low latency is critical, such as audio processing.

Formally we can describe the model as such, with $\mathcal{D}$ being the dataset, $\mathbb{N} \dots \mathbb{N}*$ being memory (how far the model can see into the past), $L$ being the number of layers, $p$ being the number of neurons per layer:
\begin{align*}
	x_i \in \mathcal{D} = x_{t-\mathbb{N}*} \dots x_{t}                                    \\
	\textit{if} \ \mathbb{N} \dots \mathbb{N} * = 3 \quad, \quad x_i = x_{t-3} \dots x_{t} \\
	\hat{y} = \Bigg[\sum^{(L,\mathbf{p})}_0\Bigg]_{\textit{GRU}}(x_{i})
\end{align*}

\subsubsection{Dataset}
The default dataset for Neural Network 3 is the Saxophone Ordinario Dataset (SOD). This dataset is consists of recordings of saxophones totaling in 90 mins of audio. Silence was removed from the dataset. The dataset has a total of 163 audio files. With the longest being approx. 3 mins in length (held note).

\subsubsection{Example Output: Distortion}
\begin{figure}
	\centering
	\includegraphics[width=0.5\columnwidth]{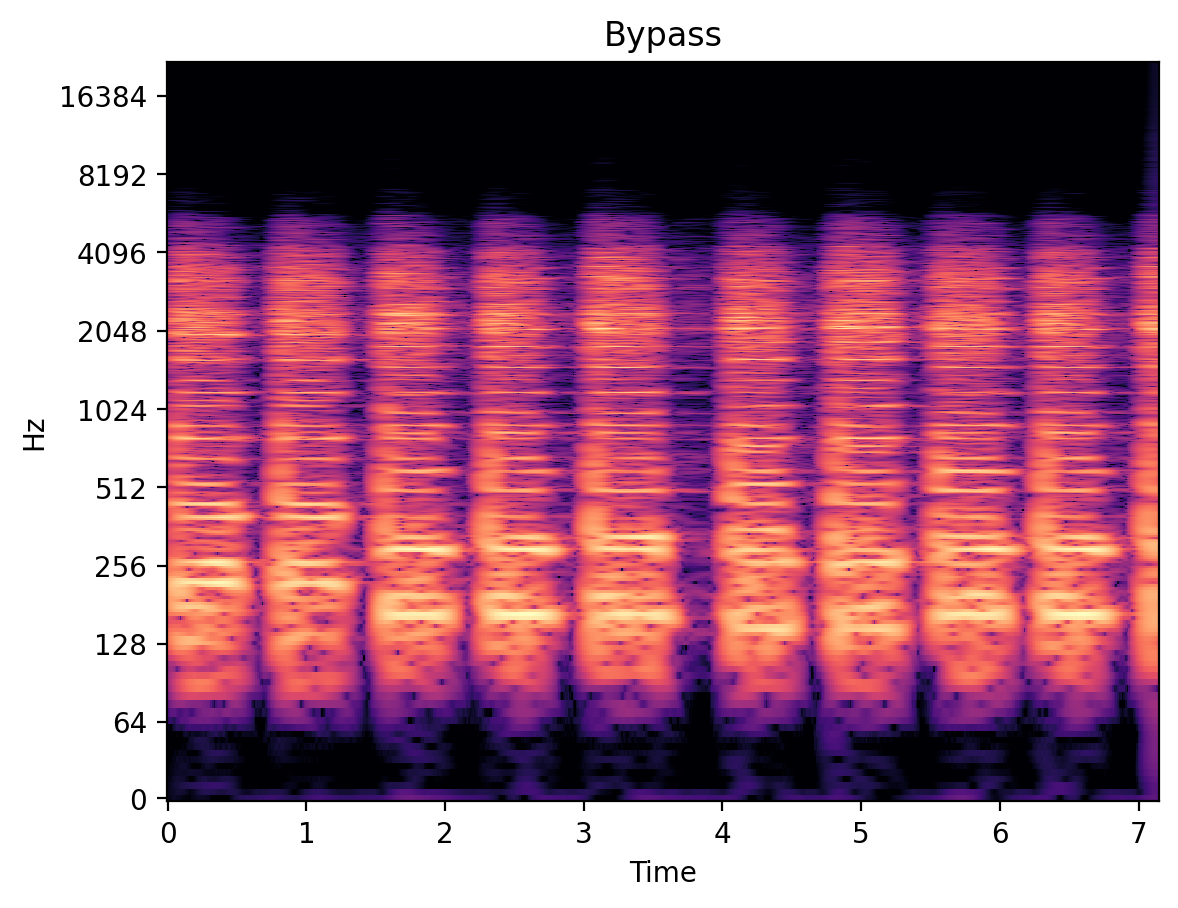}

	\caption{Distortion effect bypassed}
	\label{fig:neuralnet31}
\end{figure}

\begin{figure}
	\centering

	\includegraphics[width=0.5\columnwidth]{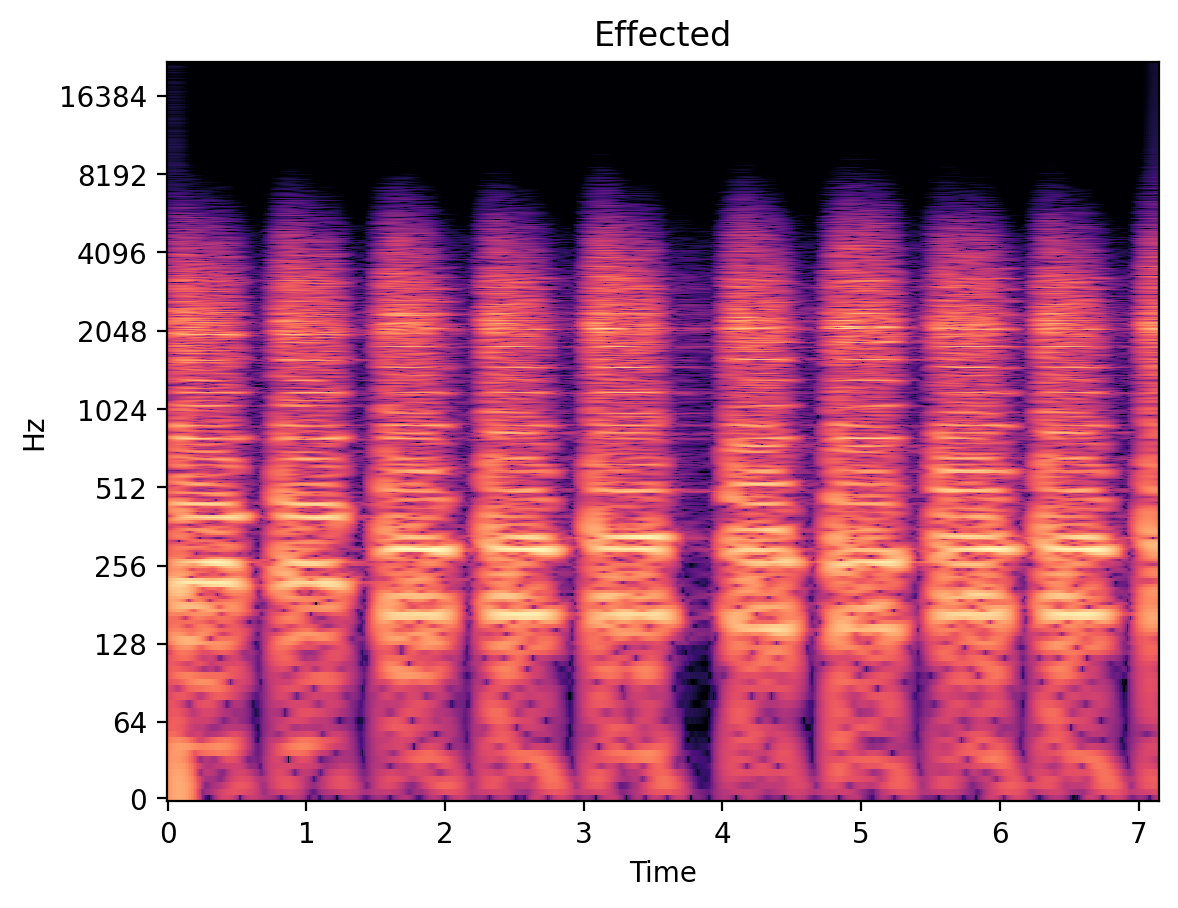}

	\caption{Distortion effect not bypassed, there is not only features of distortion above approx. 4096 Hz, but also a tail that is present. In addition to this, there is also more low frequencies around approx. < 64 Hz.}
	\label{fig:neuralnet32}
\end{figure}

The output of the result of training is a distortion effect that also contains slight reverb. For more clarification, the sound files will be present on the Github Wiki. Below shows two plots that demonstrate the distortion (shown by the presence of more harmonics) and the reverb (shown by the presence of a tail or smearing of the sound). This is shown in \autoref{fig:neuralnet31} and \autoref{fig:neuralnet32}.

\newpage
\subsubsection{Code Example}
This section demonstrates using the default code to generate a plugin for this process. The user is able to select layer count $L$, memory (which in this case would also be the number of inputs to the model), and activation function type:
\begin{lstlisting}
	activations_list = ["tanh", "relu", "sigmoid", "linear"]
	activation = activations_list[0]
	# change this from 0 - 3 to change the activation function
	MEMORY = 8
	model = Creator().createModelForNeuralNet3( 4, 
                                                    MEMORY, 
                                                    activation=activation, 
                                                    scaler=2)
\end{lstlisting}
This block of code creates Neural Network 3 with 4 layers of GRUs, 8 memory inputs, and a tanh activation function. The scaler is set to 2, which means that the model will have 2 times the number of parameters as the default model. This is useful for increasing the complexity of the model, and thus the complexity of the sounds it can generate.

\subsection{Neural Network 4: Instrument Creation (Wavetable Synthesis)}
\subsubsection{Introduction to Wavetable Synthesis}
Wavetable synthesis is a technique used in digital audio synthesis to generate complex waveforms by playing back stored samples of single-cycle waveforms. The stored samples are known as wavetables. It has been a popular and powerful method in the field of music synthesis since its inception in the 1980s. Wavetable synthesis is based on the principle of oscillators, which produce periodic signals at specific frequencies. In wavetable synthesis, these oscillators are replaced by waveforms that are stored in a table, which can be played back at different frequencies to produce different pitches. \citep{smith1991viewpoints}

The basic concept behind wavetable synthesis is to use a set of single-cycle waveforms, each representing a particular harmonic content, to create a more complex waveform that can be used to produce a musical sound. Wavetable synthesis has been used in a wide range of musical applications, including the creation of realistic acoustic instruments and the synthesis of electronic sounds. The technique has also been used in the creation of virtual analog synthesizers, which attempt to emulate the sound of traditional analog synthesizers using digital techniques.

In this work, MRCV provides a method that we are going to called Neural-generated Wavetable Synthesis. The basic concept is that the model is given a dataset of audio data, and the model is expected to produce a wavetable (user-defined size) that produces a similar matching frequency response. This is done by using a loss function that calculates the difference between the frequency response of the wavetable and the frequency response of the audio data. The model is then trained to minimize this loss function.

\subsubsection{Model: Caveat}
At the time of writing, this paper, this model is not stable and is prone to having error values that tend to infinity. However, this section will still explain the main theory behind the model. Until the bugs are fixed, we will deploy a simple feed-forward Dense network that is able to generate a wavetable based on an input of Mel-spectrograms. The currently deployed model is as such:

\begin{figure*}[h]
	\centering
	\[\begin{tikzcd} [scale cd=1]
			{Audio Data} \\
			{STFT/MFCC} & {outputsize=\{ m \times n \}} \\
			Reshape & {desiredsize = \{ 1 \times mn \}} \\
			Dense \\
			\vdots \\
			Dense \\
			GeneratedWavetable
			\arrow[from=1-1, to=2-1]
			\arrow[from=2-1, to=3-1]
			\arrow[from=3-1, to=4-1]
			\arrow[from=6-1, to=7-1]
			\arrow[dashed, no head, from=2-1, to=2-2]
			\arrow[dashed, no head, from=3-1, to=3-2]
		\end{tikzcd}\]
	\caption{The structure of the deployed Neural Network 4.}
	\label{fig:nn4_pre}
\end{figure*}
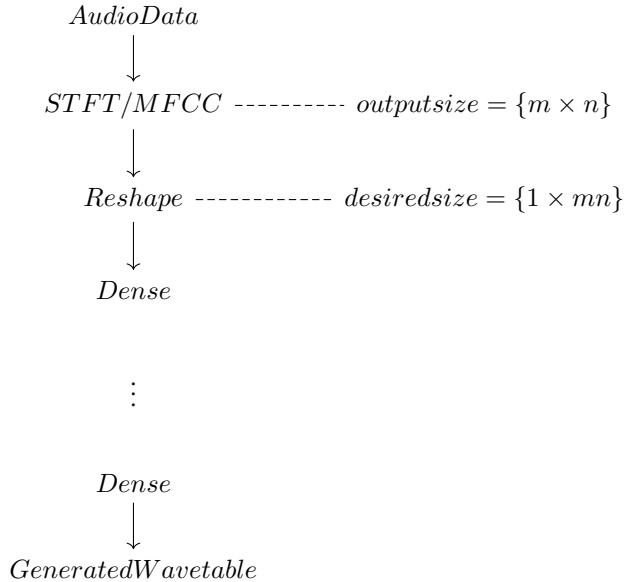

Ultimately, the model should be structurally more similar to this:
\begin{equation}
	\hat{y} = \Bigg[\sum^{(L,\mathbf{p})}_0\Bigg](x_{i})
\end{equation}
Where $x_i$ is an audio file. And in the loss function:
\begin{equation}
	\mathcal{L} = \textit{MFCC/STFT}(\hat{y})- \textit{MFCC/STFT}(x_i)
\end{equation}
It is most likely that the length of vector $\hat{y}$ $\neq$ the length of $x_i$. As $\hat{y}$ is usually around 1024 samples (as it is meant to be a wavetable, 1 single periodic oscillation of a waveform), and $x_i$ is usually more than that.

\subsubsection{Dataset}
The default dataset deployed with this model is the Piano Dataset, which is a collection of recordings of extended piano techniques \citet{pace1996lachenmann}, and ordinario piano playing. It contains 875 PCM audio files recorded at 44.1 kHz sampling rate. The dataset was cleaned up so that it does not contain any silences.

\subsubsection{Example Output}
This section will show the output of the deployed model. The model is trained on the Piano Dataset, and the output is a wavetable of size 1024. The output will then be used to procedurally generate a software synthesiser using Decent Sampler \citep{Shopdece2:online}. The output is as such:

\begin{figure}[h]
	\centering
	\includegraphics[width=0.8\columnwidth]{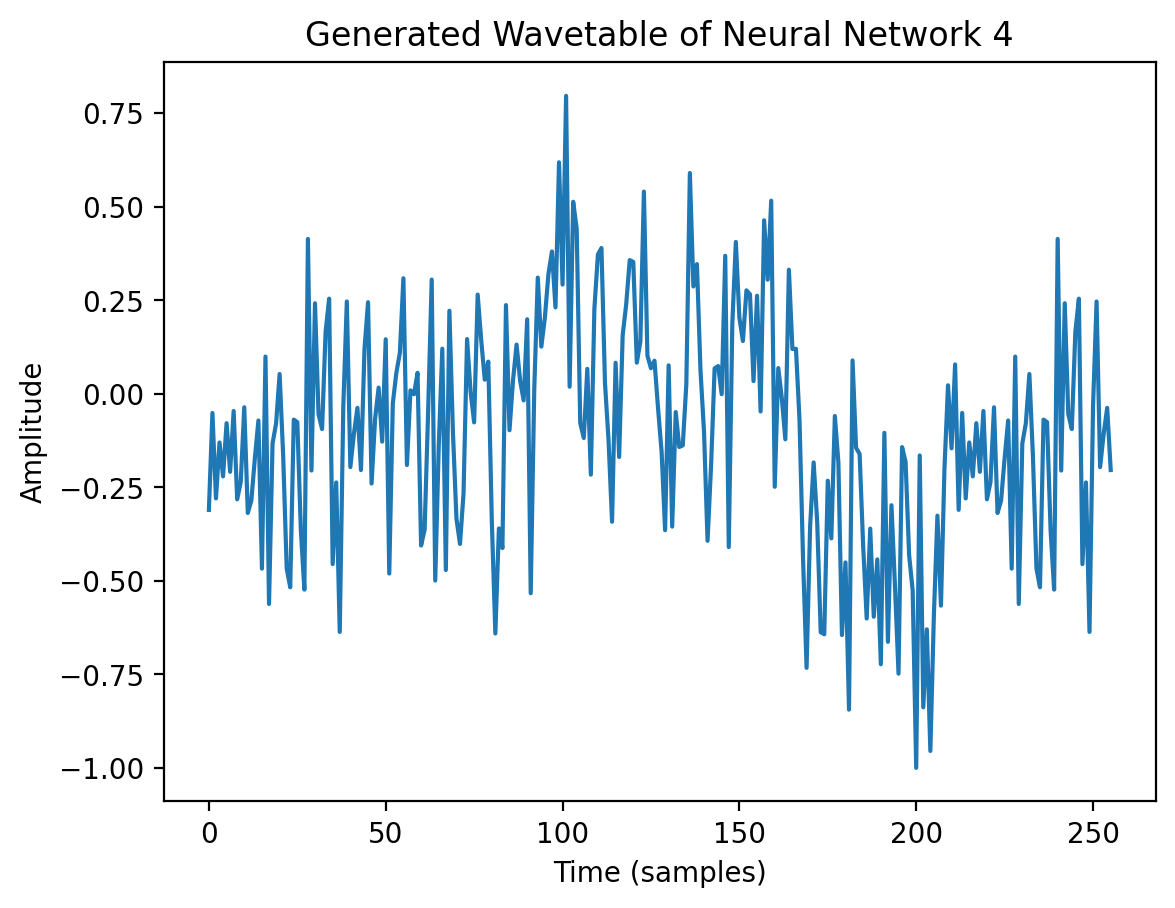}
	\caption{The result of this Neural Generated Wavetable Synthesis technique.}
\end{figure}

\subsubsection{Code Example}
The code for this neural network is similar to the preceeding examples. Users are able to define layer count $L$, layer size $p$, and block size $b$ for the model. The block size of the model determines how many samples the resulting wavetable should be:

\begin{lstlisting}
	model = Creator().createDenseModelForNeuralNet4
                            (n_layers, layer_size, input_size, BLOCK_SIZE)
\end{lstlisting}

\subsection{Genere}
Genere is a procedural midpoint between using art-centred visual editing programs (such as Adobe Illustrator/Photoshop/Indesign) and conventional notation softwares such as Sibelius/Finale/Dorico. Genere is designed to be similar to Lilypond \cite{nienhuys2003lilypond}, in that it is a code-centric environment for music notation. It differs in that the features that it provides are geared towards graphic score generation. Genere is designed to be a tool for composers to quickly generate graphic scores, and to be able to quickly edit them. Genere is also designed to be able to generate graphic scores that are not possible to be generated by hand, such as scores with thousands of notes, or scores with complex shapes. An example of this is to have each odd numbered page have larger margin sizes. While that may be a questionably hard task to do by hand or by conventional methods, it is a trivial task for Genere.

Here are two scores generated by Genere, and the code used to generate them:
\begin{figure}[h]
	\centering
	\includegraphics[width=0.8\columnwidth]{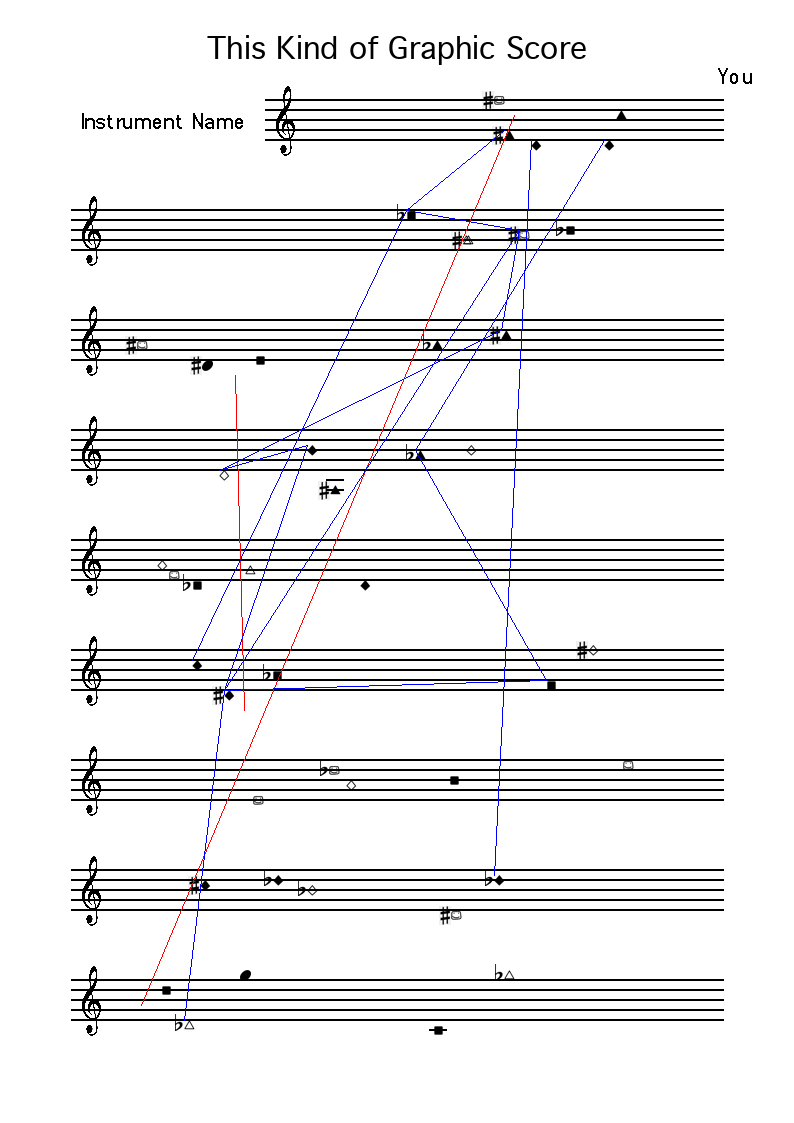}
	\caption{A showcase of some currently available features in Genere. The ability to change noteheads, draw lines, etc.}
\end{figure}
\newpage
The code used for this is as such:
\begin{lstlisting}
	GLOBAL_SAVENAME = "generic.png"
	canvas, staffLineCoords = createcanvas.returnCanvas(
	"A4", "portrait", saveRawCanvas=False, indentation=True
	)
	noter = notationplacer.notationPlacer(canvas, staffLineCoords)
	canvas = noter.applyTrebleClef(canvas, 0, on_all=True)
	canvas = noter.addTitle(canvas, "This Kind of Graphic Score")
	canvas = noter.addComposer(
	canvas, "You", pageAlignValue=4.75
	)
	canvas = noter.addInstrumentTextAtIndent(
	canvas, "Instrument Name   ", staffLineCoords, 0
	)

	notes = np.random.randint(58, 80, size=45)
	sharpOrFlat = np.random.choice(["sharp", "flat"], size=45)
	horzPositions = np.random.random_sample(size=45) * 0.8 + 0.1
	horzPositions = np.sort(horzPositions.reshape(9, 5), axis=1)
	arrayOfSystemNumbers = np.repeat(np.arange(0, 9), 5)


	for i in range(len(notes)):
	canvas = noter.applyNoteheadAt(
	canvas,
	arrayOfSystemNumbers[i],
	horzPositions[arrayOfSystemNumbers[i], i % 5],
	int(notes[i]),
	sharp_or_flat=sharpOrFlat[i],
	notehead_type=np.random.choice([...]),
	)

	dictionaryOfPlacedNotes, numberOfPlacedNotes = noter.getPlacedNotes()
	...
	canvas = noter.drawLineAcrossMultipleNotes(
	canvas, arrayOfNotes, 
        dictionaryOfPlacedNotes, 
        color="blue", line_width=1
	)
	...

	canvas.save(GLOBAL_SAVENAME)
	print(f"{GLOBAL_SAVENAME} saved to disk.")
\end{lstlisting}

\newpage
Genere also comes with a trainable Markov Model. Here is an example of the output of the Markov Model using Genere:
\begin{figure}[h]
	\centering
	\includegraphics[width=0.8\columnwidth]{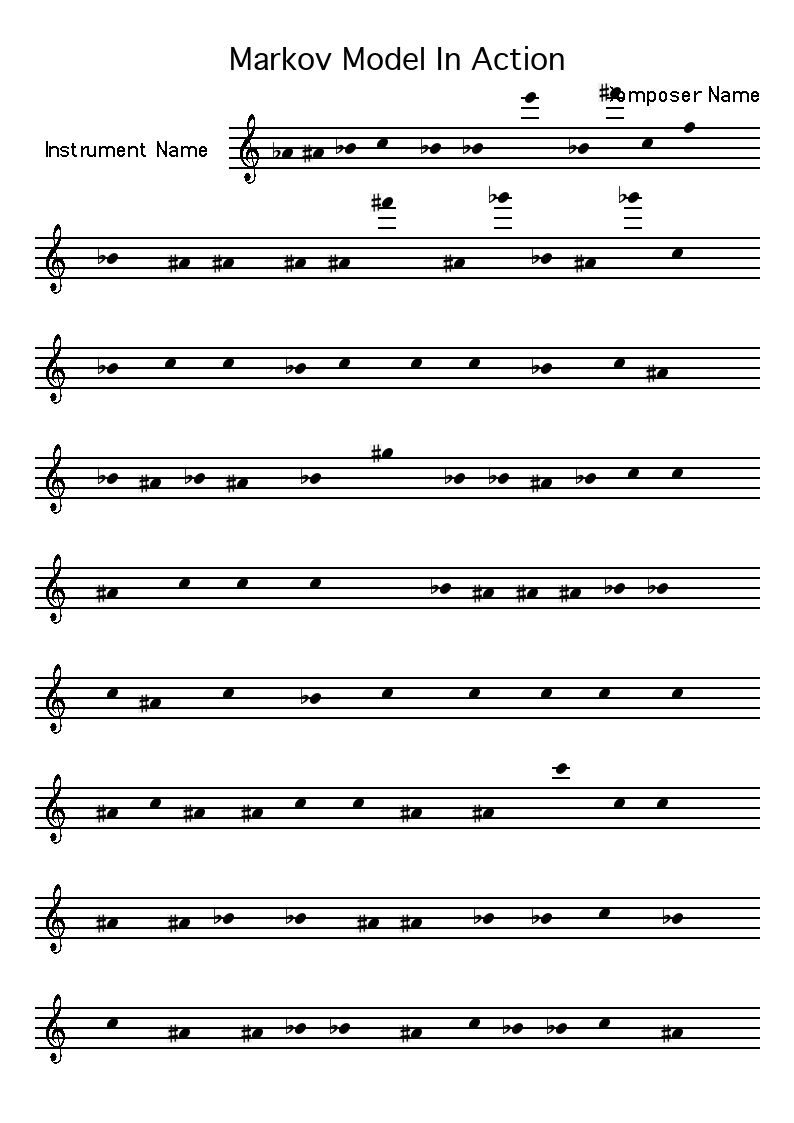}
	\caption{A showcase of the Markov Model feature in Genere.}
\end{figure}
\newpage
The code used to generate this is as such:

\begin{lstlisting}
GLOBAL_SAVENAME = "markovmodel_score.png"
canvas, staffLineCoords = createcanvas.returnCanvas(
"A4", "portrait", saveRawCanvas=False, indentation=True
)
noter = notationplacer.notationPlacer(canvas, staffLineCoords)
canvas = noter.applyTrebleClef(canvas, 0, on_all=True)
canvas = noter.addTitle(canvas, "Markov Model In Action")
canvas = noter.addComposer(canvas, "Composer Name", pageAlignValue=4.75)
canvas = noter.addInstrumentTextAtIndent
            (canvas, "Instrument Name", staffLineCoords, 0)
markov = markovmodels.MarkovModel()
rhythms, pitch, abspitch, octaves = markov.learnFromMidi(
"frenchsuite.midi"
)
# markov.plotMatrices()  # uncomment this to see the matrices
...
outputPitch, outputRhythm, outputOctave = markov.returnNextPitch()
...
canvas.save(GLOBAL_SAVENAME)
print(f"{GLOBAL_SAVENAME} saved to disk.")
\end{lstlisting}

\section{Future Work and Conclusions}
As shown above, MRCV is a software tool that uses AI and ML to help users create music, sounds, and virtual instruments. It is designed to be user-friendly for people with different levels of experience, and its main goal is to encourage creativity. MRCV allows users to customize input datasets and offers many options for each neural network. The documentation is designed to be easy to understand and abstracts technical details. MRCV is open source, meaning users can contribute to its development and the community can benefit from each other's experience.

For future work on the system, users can:
\begin{itemize}
	\item Add more neural networks in the form of trained models (in \texttt{.h5} or \texttt{.json} (for realtime audio) format)
	\item Add more features to Genere (such as the ability to generate a vector image, it is currently a bitmap)
	\item Add more features via requests on the GitHub page or answer feature requests.
	\item Spread the word to other people who may be interested in using MRCV, as an introduction to AI and ML in music.
\end{itemize}
\newpage
\bibliographystyle{unsrtnat}
\bibliography{references}

\end{document}